\documentclass[10pt,twocolumn]{article}
\usepackage{amsmath}
\usepackage{amssymb}
\usepackage{epsfig}

\begin{document}

\setlength{\baselineskip}{12pt}

\title{Energy-aware Load Balancing Policies for the Cloud Ecosystem}
\author{Ashkan Paya and Dan C. Marinescu \\
Computer Science Division \\
Department of Electrical Engineering and Computer Science \\
University of Central Florida, Orlando, FL 32816, USA \\
Email:ashkan\_paya@knights.ucf.edu, dcm@cs.ucf.edu}

\maketitle

\begin{abstract}
The energy consumption of computer and communication systems does not scale linearly with the workload. A system uses a significant amount of energy even when idle or lightly loaded. A widely  reported solution to resource management in large data centers is to concentrate the load on a subset of servers  and, whenever possible, switch the rest of the servers to one of the possible sleep states. We propose a reformulation of the traditional concept of load balancing aiming to optimize the energy consumption of a large-scale system:  {\it distribute the workload evenly to the smallest set of servers operating at an optimal energy level, while observing QoS constraints, such as the response time.} Our model applies to clustered systems; the model also requires that the demand for system resources to increase at a bounded rate in each reallocation interval. In this paper we report the VM migration costs for application scaling.
\end{abstract}

\section{Introduction and Motivation}
\label{Introduction}
\medskip

The concept of ``load balancing'' dates back to the time the first distributed computing systems
were implemented in the late 1970s and early 1980s. It means exactly what the name implies, {\it to evenly distribute the workload to a set of servers} to maximize the throughput, minimize the response time, and increase the system resilience to faults by avoiding overloading one  or more systems in the distributed environment.

Distributed systems became popular after communication networks allowed multiple computing engines to effectively communicate with one another and the networking software became an integral component of an operating system. Once processes were able to easily communicate with one another using sockets\footnote{ The sockets were introduced by BSD (Berkeley Systems Distribution) Unix in 1977}, the {\it client-server paradigm} became the preferred method to develop distributed applications; it enforces modularity, provides a complete isolation of clients from the servers, and enables the development of stateless servers.

The client-server model proved to be not only enduring, but also increasingly successful; three decades later, it is at the heart of utility computing. In the last few years packaging computing cycles and storage and offering them as a metered service became a reality. Large farms of computing and storage platforms have been assembled and a fair number of Cloud Service Providers (CSPs) offer computing and storage services based on three different delivery models SaaS (Software as a Service), PaaS (Platform as a Service), and IaaS (Infrastructure as a Service).

Reduction of energy consumption thus, of the carbon footprint of cloud related activities, is increasingly more important for the society. Indeed, as more and more applications run on clouds, more energy is required to support cloud computing than the energy required for many other human-related activities. While most of the energy used by data centers is directly  related to cloud computing, a significant fraction is also used by the networking infrastructure used to access the cloud. This fraction is increasing, as wireless access becomes more popular and wireless communication is energy intensive. In this paper we are only concerned with a single aspect of energy optimization, minimizing the energy used by cloud servers.

Unfortunately, computer and communication systems are not energy proportional systems, in other words, their energy consumption does not scale linearly with the workload; an idle system consumes a rather significant fraction, often as much as $50\%$, of the energy used to deliver peak performance.
Cloud elasticity, one of the main attractions for cloud users, comes at a stiff price as the cloud resource management is based on {\it over-provisioning}. This means that a cloud service provider has to invest in a larger infrastructure than a ``typical'' or average cloud load warrants. At the same time, cloud elasticity implies that most of the time cloud servers operate with a low load, but still use a large fraction of the energy necessary to deliver peak performance. The low average cloud server utilization  \cite{Abts10,Ardagna11,Google13,Marinescu13} affects negatively the common measure of energy efficiency, the {\it performance per Watt of power} and amplifies the ecological impact of cloud computing.

The strategy for resource management in a computing cloud we discuss is to concentrate the load on a subset of servers  and, whenever possible, switch the rest of the servers to a sleep state. In a sleep state the energy consumption is very low. This observation implies that the traditional concept of load balancing could be reformulated  to optimize the energy consumption of a large-scale system as follows: {\it distribute evenly the workload to the smallest set of servers operating at an optimal energy level, while observing QoS constraints, such as the response time.} An {\it optimal energy level} is one when the  {\it normalized system performance}, defined as the ratio of the current performance to the maximum performance, is delivered with the minimum {\it normalized energy consumption}, defined as the ratio of the current energy consumption to the maximal one.

From the large number of questions posed by energy-aware load balancing policies discussed in Section \ref{EnergyOptimization}, we discuss only the energy costs for migrating a VM when we decide to either switch a server to a sleep state or force it to operate within the boundaries of an energy optimal regime.

\section{Operating Efficiency of a System}
\label{EnergyProportionalSystems}
\medskip
In this section we discuss energy proportional systems, the dynamic range of different subsystems  including memory, secondary storage, and interconnection networks. Then we overview methods to reduce the energy consumption of a computer system and discuss sleep states.
 
{\bf Operating efficiency.} The operating efficiency of a system is captured by an expression of ``performance per Watt of power.'' It is reported that during the last two decades the performance of computing systems has increased much faster than their operating efficiency; for example, during the period $1998$ till $2007$, the performance of supercomputers has increased $7,000\%$  while their operating efficiency has increased only $2,000\%$. Recall that power is the amount of energy consumed per unit of time and it is measured in Watts, or Joules/second.

{\bf Energy proportional systems.} In an ideal world, the energy consumed by an idle system should be near zero and grow linearly with the system load. In real life, even systems whose power requirements scale linearly,  when idle use more than half the power they use at full load. Data collected over a long period of time shows that the typical operating region for data center servers is far from an optimal energy consumption region as we shall see in Section \ref{EnergyOptimization}.

Energy-proportional systems could lead to large savings in energy costs for computing clouds.
An {\it energy-proportional} system consumes no power when idle, very little power under a light load and, gradually, more power as the load increases. By definition, an ideal energy-proportional system is always operating at $100\%$ efficiency. Humans are a good approximation of an ideal energy proportional system; the human energy consumption is about $70$ W at rest, $120$ W on average on a daily basis, and can go as high as $1,000 - 2,000$ W during a strenuous, short time effort \cite{Barroso07}.

{\bf Dynamic range of subsystems.} The dynamic range is the difference between the upper and the lower limits of the energy consumption of a system function of the load placed on the system. A large dynamic range means that a system is able to operate at a lower fraction of its peak energy when its load is low. Different subsystems of a computing system behave differently in terms of energy efficiency; while many processors have reasonably good energy-proportional profiles,  significant improvements in memory and disk subsystems are necessary. 

The processors used in servers consume less than one-third of their peak power at very-low load and have a dynamic range of more than $70\%$ of peak power; the processors used in mobile and/or embedded applications are better in this respect. According to \cite{Barroso07} the dynamic power range of other components of a system  is much narrower: less than $50\%$ for DRAM, $25\%$ for disk drives, and $15\%$ for networking switches.

The largest consumer of power of a system is the processor, followed by memory, and storage systems. The power consumption can vary from 45W to 200W per multi-core CPU; newer processors include power saving technologies. Large servers often use $32-64$ dual in-line memory modules (DIMMs); the power consumption of one DIMM is in the $5 - 21$ W range.  Server secondary memory cooling requires additional power; a server with $2-4$ hard disk drives (HDDs) consumes $24 - 48$ W.

A strategy to reduce energy consumption by disk drives is to concentrate the workload on a small number of disks and allow the others to operate in a low-power mode. One of the techniques to accomplish this  is based on replication. A replication strategy based on a sliding window is reported in \cite{Vrbsky10};  measurement results indicate that it performs better than  LRU, MRU, and LFU\footnote{LRU (Least Recently Used), MRU (Most Recently Used), and LFU(Least Frequently Used) are replacement policies used by memory hierarchies  for caching and paging.} policies for a range of file sizes, file availability, and number of client nodes and the power requirement is reduced by as much as $31\%$.

Another technique is based on data migration. The system in \cite{Hasebe10} uses data storage in virtual nodes managed with a distributed hash table; the migration is controlled by two algorithms,  a short-term optimization algorithm used for gathering or spreading virtual nodes according to the daily variation of the workload so that the number of active physical nodes is reduced to a minimum, and a  long-term optimization algorithm, used for coping with changes in the popularity of data over a longer period, e.g., a week.

A number of proposals have emerged for {\it energy proportional} networks; the energy consumed by such networks is proportional with the communication load. For example, in \cite{Abts10} the authors argue that a data center network based on a flattened butterfly topology is more energy and cost efficient. High-speed channels typically consist of multiple serial lanes with the same data rate;  a physical unit is stripped across all the active lanes. Channels commonly operate plesiochronously\footnote{Different parts of the system are almost, but not quite perfectly, synchronized; in this case, the core logic in the router operates at a frequency different from  that of the I/O channels.} and are always on, regardless of the load, because they must still send idle packets to maintain byte and line alignment across the multiple lines. An example of an energy proportional network is {\it InfiniBand}.

\begin{table*}[!ht]
\begin{center}
\caption{Estimated average power use of volume, mid-range, and high-end servers (in Watts) along the years \cite{Koomey07}.}
\label{ServerEnergyConsumption}
\begin{tabular} {|c|ccccccc|}
\hline
Type   & 2000  & 2001  & 2002  & 2003  & 2004  & 2005  & 2006 \\
\hline
Vol    & 186   & 193   & 200   & 207   & 213   & 219   & 225 \\
Mid    & 424   & 457   & 491   & 524   & 574   & 625   & 675 \\
High   & 5,534 & 5,832 & 6,130 & 6,428 & 6,973 & 7,651 & 8,163 \\
 \hline
\end{tabular}
\end{center}
\end{table*}

{\bf Sleep states.} A comprehensive document \cite{ACPI11} elaborated by Hewlett-Packard, Intel, Microsoft, Phoenix Technologies, and Toshiba describes the advanced configuration and power interface (ACPI) specifications which allow an operating system (OS) to effectively manage the power consumption of the hardware. Several types of {\it sleep sates}, are defined: C-states (C1-C6) for the CPU, D-states (D0-D3) for modems, hard-drives, and CD-ROM, and S-states (S1-S4) for the basic input-output system (BIOS).

The C-states, allow a computer to save energy when the CPU is idle. In a sleep state the idle units of a CPU have their clock signal and the power cut.  The higher the state number, the deeper the CPU sleep mode, the larger the energy saved, and the longer the time for the CPU to return to the state $C0$ which corresponds to the CPU fully operational. In states $C1$ to $C3$ the clock signal and the power of different CPU units are cut, while in states $C4$ to $C6$ the CPU voltage is reduced. For example, in the C1 state the main internal CPU clock is stopped by the software but the bus interface is and the advanced programmable interrupt controller (APIC) are running, while in state C3 all internal clocks are stopped, and in state C4 the CPU voltage is reduced.

{\bf Economy of scale.} Economy of scale affects the energy efficiency of data processing \cite{Gandhi11}. For example, Google reports that the annual energy consumption for an Email service varies significantly depending on the business size and can be 15 times larger for a small one \cite{Google13}. Cloud computing can be more energy efficient than on-premise computing for many organizations \cite{Baliga11, NRDC12}.

The power consumption of servers has increased over time. Table \ref{ServerEnergyConsumption}  \cite{Koomey07} shows the evolution of the average power consumption for volume (Vol) servers - servers with a price less than \$ 25 K, mid-range (Mid) servers - servers with a price between \$25 K and \$499 K, and high-end (High) servers - servers with a price tag larger than \$500 K.

The energy to transport data is a significant component of the total energy cost. According to \cite{Baliga11} ''a public cloud could consume three to four times more power than a private one due to increased energy consumption in transport.''

\section{Energy optimization in large-scale data centers}
\label{EnergyOptimization}
\medskip

{\bf Motivation.} Recently, Gartner research reported that the average server utilization in large data-centers is $18\%$ \cite{Snyder10}, while the utilization of {\it x86} servers is even lower, $12\%$. These results confirm earlier estimations that the average server utilization is in the $10 - 30\%$ range  \cite{Barroso07}. A 2010 survey \cite{Blackburn10} reports that idle servers contribute $11$ million tones of unnecessary $CO_{2}$ emissions each year and that the total yearly costs for idle servers is $\$19$ billion.

The alternative to the wasteful resource management policy when the servers are {\it always on}, regardless of their load, is to develop {\it energy-aware load balancing} policies. Such policies combine {\it dynamic power management} with load balancing and attempt to identify servers operating outside their optimal power regime and decide if and when they should be switched to a sleep state or what other actions should be taken to optimize the energy consumption.  The term {\it server consolidation} is sometimes used to describe the process of switching idle systems to a sleep state.

{\bf Challenges and metrics for energy-aware load balancing.} Some of the questions posed by energy-aware load balancing are: 

\begin{enumerate}

\item Under what conditions should a server be switched to a sleep state? 

\item  What sleep state should the server be switched to?

\item How much energy is necessary to switch a server to a sleep state and then switch it back to an active state?

\item How much time it takes to switch a server in a sleep state to a running state? 

\item How much energy is necessary to migrate a VM running on a server to another one? 

\item How much energy is necessary to start a VM on the target server? 

\item  How to choose the target for the migration of a  VM? 

\item How much time it takes to migrate a VM?

\end{enumerate}

Two basic metrics ultimately determine the quality of an energy-aware load balancing policy: (1) the amount of energy saved; and (2) the number of violations it causes. In practice, the metrics depend on the system load and other resource management policies, e.g., the admission control policy and the QoS guarantees offered. The load can be slow- or fast-varying, have spikes or be smooth, can be predicted or is totally unpredictable; the admission control can restrict the acceptance of additional load when the available capacity of the servers is low. What we can measure in practice is the {\it average energy} used and the average server {\it setup time}. The setup time varies depending on the hardware and the operating system and can be as large as $260$ seconds \cite{Gandhi12}; the energy consumption during the setup phase is close the maximal one for the server.

The time to switch the servers to a running state is critical when the load is fast varying, the load variations are very steep, and the spikes are unpredictable. The decisions when to switch servers to a sleep state and back to a running state are less critical when a strict admission control policy is in place;  then new service requests for large amounts of resources can be delayed until the system is able to turn on a number of sleeping servers to satisfy the additional demand.

\begin{figure*}[!ht]
\begin{center}
\includegraphics[width=12cm]{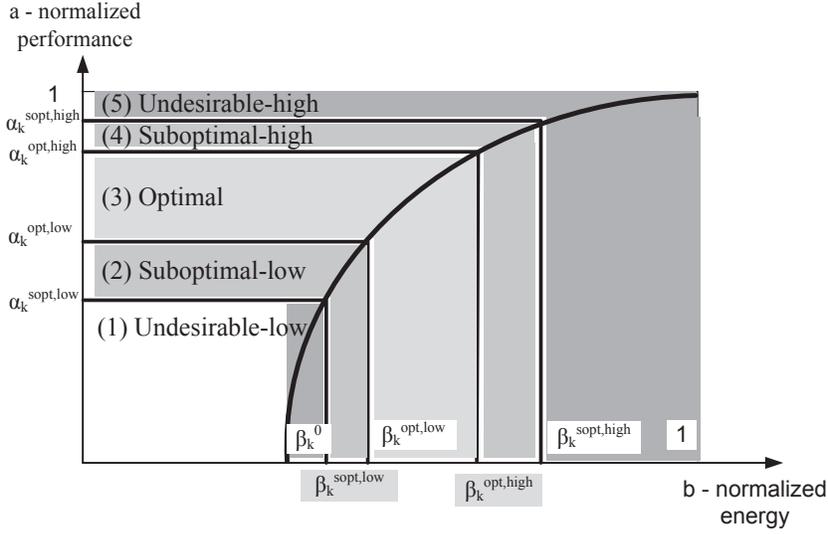}
\end{center}
\caption{Normalized performance versus normalized server energy consumption; the boundaries of the five  operating regions are shown.}
\label{ServerPerformanceVsEnergy}
\end{figure*}

{\bf Policies and mechanisms to implement the policies.} Several policies have been proposed to decide when to switch a server to a sleep state. The {\it reactive} policy \cite{Urgaonkar05} responds to the current load, it switches the servers to a sleep state when the load decreases and switches them to the running state when the load increases. Generally, this policy leads to SLA violations and could work only for slowly-varying and predictable loads. To reduce SLA violations one can envision a {\it reactive with extra capacity} policy when one attempts to have a safety margin and keep a fraction of the total number of servers, e.g., $20\%$, running above those needed for the current load. The {\it autoscale} policy \cite{Gandhi12} is a very conservative {\it reactive} policy in switching servers to sleep state to avoid the power consumption and the delay in switching them back to running state. This can be advantageous for unpredictable, spiky loads.

A very different approach is taken by two version of {\it predictive} policies \cite{Bodik09,Verma09}. The {\it moving window averages} one estimates the workload by measuring the average request rate in a window of size $\Delta$ seconds and use this average to predict the load during the next second (second $\Delta+1$) and then slide the window one second to predict the load for second $\Delta+2$, and so on. The {\it predictive linear regression} policy uses a linear regression to predict the future load.

An {\it optimal} policy can be defined as one which does not produce any SLA violations and guarantees that all servers operate in their optimal energy regime. Optimality is a local property of a server and can be easily determined by the energy management component of the hypervisor. Recall that $E_{k}^{opt}$, the optimal energy level of server $\mathcal{S}_{k}$  is one when the {\it normalized system performance}, defined as the ratio of the current performance to the maximum performance is delivered with the minimum {normalized energy consumption}, defined as the ratio of the current energy consumption to the maximal one. The boundaries of the optimal regions are defined as $E_{k}^{opt} \pm \delta$ with $\delta=(0.05 - 0.1) \times E_{k}^{opt}$. In a heterogeneous environment the normalized system performance and the normalized energy consumption differ from server to server. 

The mechanisms to implement energy-aware load balancing policies should  satisfy several conditions: 

\begin{enumerate}
\item Scalability - work well for large farms of servers. 

\item Effectiveness  - lead to substantial energy and cost savings. 

\item Practicality - use efficient algorithms and require as input only data that can be measured with low overhead and, at the same time, accurately reflects the state of the system.

\item  Consistency -  the policies should be aligned with the global system objectives and with the contactual obligations specified by Service Level Agreements, e.g., observe deadlines, minimize the response time, and so on.
\end{enumerate}

\section{Clustered Cloud Models}
\label{SimpleModel}
\medskip

{\bf Clustering.} Hierarchical organization  has long been recognized as an effective way to cope with system complexity.  Clustering supports scalability,  as the number of systems increase we add new clusters. Clustering also supports practicality, server decisions are based primarily on local state information gathered from the members of the cluster; such information is more accurate and available with lower overhead then information from a very large population. 

{\bf An energy-aware model.} In  \cite{Paya13} we introduced a model of large-scale system with a clustered organization. In this model we distinguish several regimes of operation for a server based on the energy efficiency. 

We assume that the normalized performance of server $\mathcal{S}_{k}$ depends on the energy level $a_{k}(t) = f_{k} [b_{k}(t)]$ and  distinguish five operating regions of a server, an optimal one, two sub-optimal, and two undesirable, Figure \ref{ServerPerformanceVsEnergy}:
\smallskip

$\mathcal{R}_1$ - undesirable low region
\begin{equation}
\label{un-low}
\beta_{k}^{0}  \le b_{k}(t) \le \beta_{k}^{sopt,l} ~~~0  \le {a_{k}(t)} \le \alpha_{k}^{sopt,l}
\end{equation}

$\mathcal{R}_{2}$ - lower suboptimal region
\begin{equation}
\label{sob-low}
\beta_{k}^{sopt,l}  \le b_{k}(t) \le \beta_{k}^{opt,l} ~~~
\alpha_{k}^{sopt,l}  \le {a_{k}(t)} \le \alpha_{k}^{opt,l}.
\end{equation}

$\mathcal{R}_{3}$ - optimal  region
\begin{equation}
\label{optimized}
\beta_{k}^{opt,l}  \le b_{k}(t) \le \beta_{k}^{opt,h}~~~
\alpha_{k}^{opt,l}  \le { a_{k}(t)} \le \alpha_{k}^{opt,h}.
\end{equation}

$\mathcal{R}_{4}$ - upper suboptimal region
\begin{equation}
\label{sob-high}
\beta_{k}^{opt,h}  \le b_{k}(t) \le \beta_{k}^{sopt,h}~~~
\alpha_{k}^{opt,h}  \le {a_{k}(t)} \le \alpha_{k}^{sopt,h}.
\end{equation}

$\mathcal{R}_5$ - undesirable high (h) region
\begin{equation}
\label{un-high}
\beta_{k}^{sopt, h}  \le b_{k}(t) \le 1 ~~~
\alpha_{k}^{sopt,h}  \le { a_{k}(t)} \le 1
\end{equation}

The classification captures the current system load and allows us to distinguish the actions to be taken to return to the optimal regime or region. When the system is operating in the upper suboptimal or undesirable regions one or more VMs should be migrated elsewhere to lower the load the load of the server. When operating in the lower suboptimal or undesirable regions the system is lightly loaded; then additional load should be brought in, or, alternatively, the system should be a candidate for switching to a sleep state. This classification also captures the urgency of the actions taken; suboptimal regions do not require an immediate attention, while undesirable regions do. The time spent operating in each non-optimal region is also important. Of course one can further refine the model and define a larger number of regions but this could complicate the algorithms.

An important characteristic of the model is that the rate of workload increase is limited. This requirement is motivated by the fact that effective admission control policies are rarely effective because the available capacity of the a cloud is difficult to estimate; it should be a part of a Service Level Agreement.

{\bf A homogeneous cloud model.} To estimate the energy savings by distributing the work load to the smallest set of servers operating at an optimal energy mode we consider a simple model. The model assumes a homogeneous environment all servers have the same peak performance and the same energy consumption at the peak performance. Moreover, it ignores the overhead of migrating computations from one server to another during the energy optimization process. We compare the energy consumption for two scenarios:
\begin{itemize}
\item
{\it Reference cloud operation} - the $n$ physical platforms operate at normalized performance levels uniformly distributed in the interval $[a_{min}, a_{max}] $. We assume that the average normalized energy consumption per operation in this range is $b_{avg}$. Then the energy consumption is

\begin{equation}
E_{ref} = n b_{avg}
\end{equation}
and the number of operations is
\begin{equation}
C_{ref} = n a_{avg} ~~~with~~a_{avg} = {{a_{max} - a_{min}} \over {2}}
\end{equation}
\item
{\it Optimal energy-operation} - a subset, $n_{sleep} < n $ servers are switched to a sleep state and the remaining $(n-n_{sleep})$ platforms operate at a normalized performance level $a_{opt}$ and  the normalized energy consumption per operation is $b_{opt} = b_{avg} + \epsilon$. Then the energy consumption is
\begin{equation}
E_{opt} = (n - n_{sleep}) b_{opt}
\end{equation}
and the number of operations is
\begin{equation}
C_{opt} = (n - n_{sleep}) a_{opt}
\end{equation}
\end{itemize}

The ratio of energy consumption for the two scenarios is

\begin{equation}
{ E_{ref} \over E_{opt}} = {{n  \over {n - n_{sleep}}} \times {b_{avg} \over {b_{opt}}}}
\end{equation}

We require that the volume of computations carried out under the two scenarios be the same

\begin{equation}
n a_{avg} =  (n - n_{sleep}) a_{opt}~~\Rightarrow~~{ {n} \over {n - n_{sleep}} } = { {a_{opt}} \over {a_{avg}} }.
\end{equation}
It follows that

\begin{equation}
{ E_{ref} \over E_{opt}} = {a_{opt} \over a_{avg}} \times {b_{avg} \over {b_{opt}}}
\end{equation}
For example, when $b_{avg}=0.6$, $a_{avg}=0.3$,
$b_{opt}=0.8$, and $a_{opt}=0.9$  then

\begin{equation}
{ E_{ref} \over E_{opt}} = 2.25.
\end{equation}
In this case the optimal operation reduces the energy consumption to less than half.

{\bf A heterogeneous cloud model.} A more complex model is used for the second type of simulation experiments. Some of the model parameters are: $\tau_{k}$ the reallocation interval, $\lambda_{i,k}$ the largest rate of increase in demand for CPU cycles of the application $A_{i,k}$ on server $S_{k}$,  $q_{k}(t+\tau_{k})$ and $p_{k}( t+\tau_{k})$ the costs for horizontal and vertical scaling, respectively for server $S_{k}$ in the next reallocation interval, and $ j_{k}(t+\tau_{k}) $, cost of communication and data transfer to or from the leader for the next reallocation interval. The average server load is uniformly distributed in the $[0.1 - 0.9]$ range of the normalized performance. The servers are connected to the leader by star topology. 

The parameters defining the energy regimes for server $S_{k}$,  $\alpha_{k}^{sopt,l}, \alpha_{k}^{opt,l}, \alpha_{k}^{opt,h}$, and $\alpha_{k}^{sopt,h}$ are randomly chosen  from  uniform distributions in the [0.20 - 0.25], [0.25 - 0.45], [0.55 - 0.80] and [0.80 - 0.85] range, respectively. Each application has a unique $\lambda_{i,k}$. Initially, all servers are operating in the $C_{0}$ mode. A server $S_{k}$ maintains static information such as the serverId, and the boundaries of the energy regimes, $\alpha_{k}^{sopt,l}, \alpha_{k}^{opt,l}, \alpha_{k}^{opt,h}$, and $\alpha_{k}^{sopt,h}$; it also maintains dynamic information such as the number of applications, the server load, the regime of operation and CPU state. The leader is informed periodically about the regime of each server of the cluster.

At the end of the current reallocation interval, $R_{k}(t)$, server $S_{k}$ evaluates the operation regime for the next reallocation interval,  $R_{k}(t+\tau_{k})$ based on the server load, CPU cycles demand of each individual application and performance of the server. If needed, it also calculates the costs for horizontal and vertical scaling and communication with the leader which depends on the regime of operation and number of applications. In other words, $S_{k}$ computes $a_{k}(t+\tau_{k}), q_{k}(t+\tau_{k}), p_{k}( t+\tau_{k})$ and $ j_{k}(t+\tau_{k})$. After determining the regime of operation in the next reallocation interval, $R_{k}(t+\tau_{k})$. If the regime is:

\begin{enumerate}
\item
$\mathcal{R}_{1}$ - then $S_{k}$ notifies the leader. Upon receiving the notification, the searches for servers operating in the $\mathcal{R}_{4}$ or $\mathcal{R}_{5}$ regimes, as well as, other servers operating in the $\mathcal{R}_{1}$ or $\mathcal{R}_{2}$ regimes.  If such servers are identified, the leader calculates the cost of transferring VMs from/to such servers and sends this information to server $S_{k}$. Upon receiving this information, $S_{k}$ determines if it should gather additional workload from servers operating in either $\mathcal{R}_{4}$ or $\mathcal{R}_{5}$ regimes, or if it should transfer its own workload to servers  operating in the $\mathcal{R}_{1}$ or $\mathcal{R}_{2}$ regimes and then switch itself to sleep.
\item
$\mathcal{R}_{2}$ - then  $S_{k}$ notifies the leader that it is willing to accept additional workload. The leader searches for servers operating in the $\mathcal{R}_{4}$ and $\mathcal{R}_{5}$ regimes and informs them that $S_{k}$ is willing to accept some of their workload. Finally, $S_{k}$ negotiates directly with the potential partners for load balancing.
\item
$\mathcal{R}_{3}$ - no action is necessary.
\item
$\mathcal{R}_{4}$ - then $S_{k}$  notifies the leader that it is overloaded. The leader searches for servers operating in the $\mathcal{R}_{1}$ and $\mathcal{R}_{2}$ regimes and informs them that $S_{k}$ wishes to transfer some of its workload to them. Finally, $S_{k}$ negotiates directly with the potential partners for load balancing.
\item
$\mathcal{R}_{5}$ - then $S_{k}$ notifies the leader. Upon receiving the notification, the leader searches for servers operating in the $\mathcal{R}_{1}$ or $\mathcal{R}_{2}$ regimes and requests $S_{k}$ to  negotiate directly with the potential partners for load balancing. If no such servers are found, the leader wakes up one or more servers in $C_{3}$ or $C_{6}$ (sleep) states and informs $S_{k}$.
\end{enumerate}

\section{Simulation Experiments}
\label{SimulationExperiments}
\medskip

{\bf The effect of the system load.} In \cite{Paya13} we report on simulation experiments designed to evaluate the effectiveness of algorithms to balance the load while attempting to optimize the energy consumption while accepting additional load. One of the questions we addressed was whether the system load has an effect on the resource management strategy to force the servers in a cluster to operate within the boundaries of the optimal region. In \cite{Paya13} we experimented with clusters sizes 20, 40, 60, and 80 servers; the experiments we report now are for the clusters with $10^2, 10^3$, and $10^4$ servers. For each cluster size we considered two load distributions:

\begin{figure*}[!ht]
\begin{center}
\includegraphics[width=9cm]{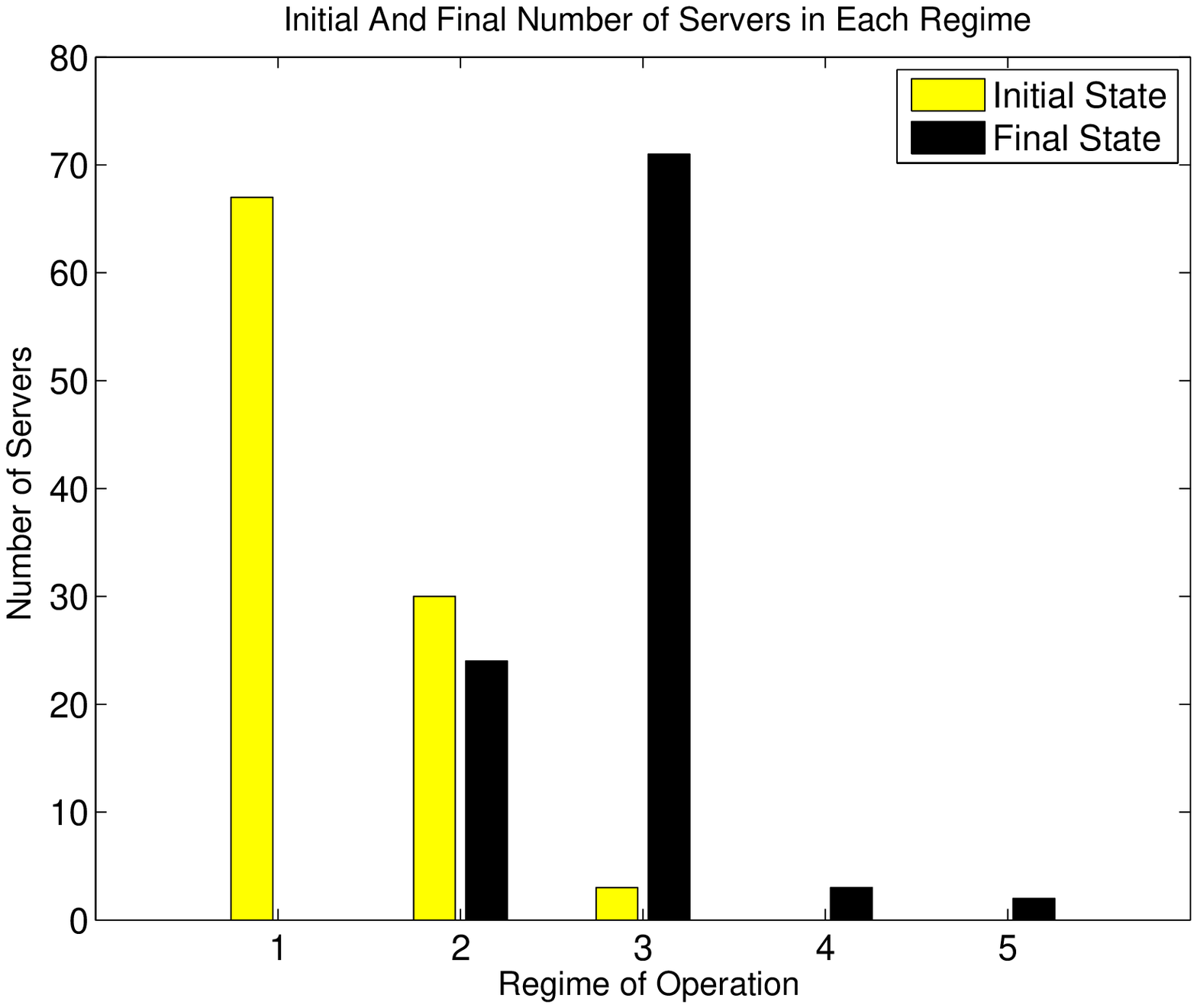}
\includegraphics[width=9cm]{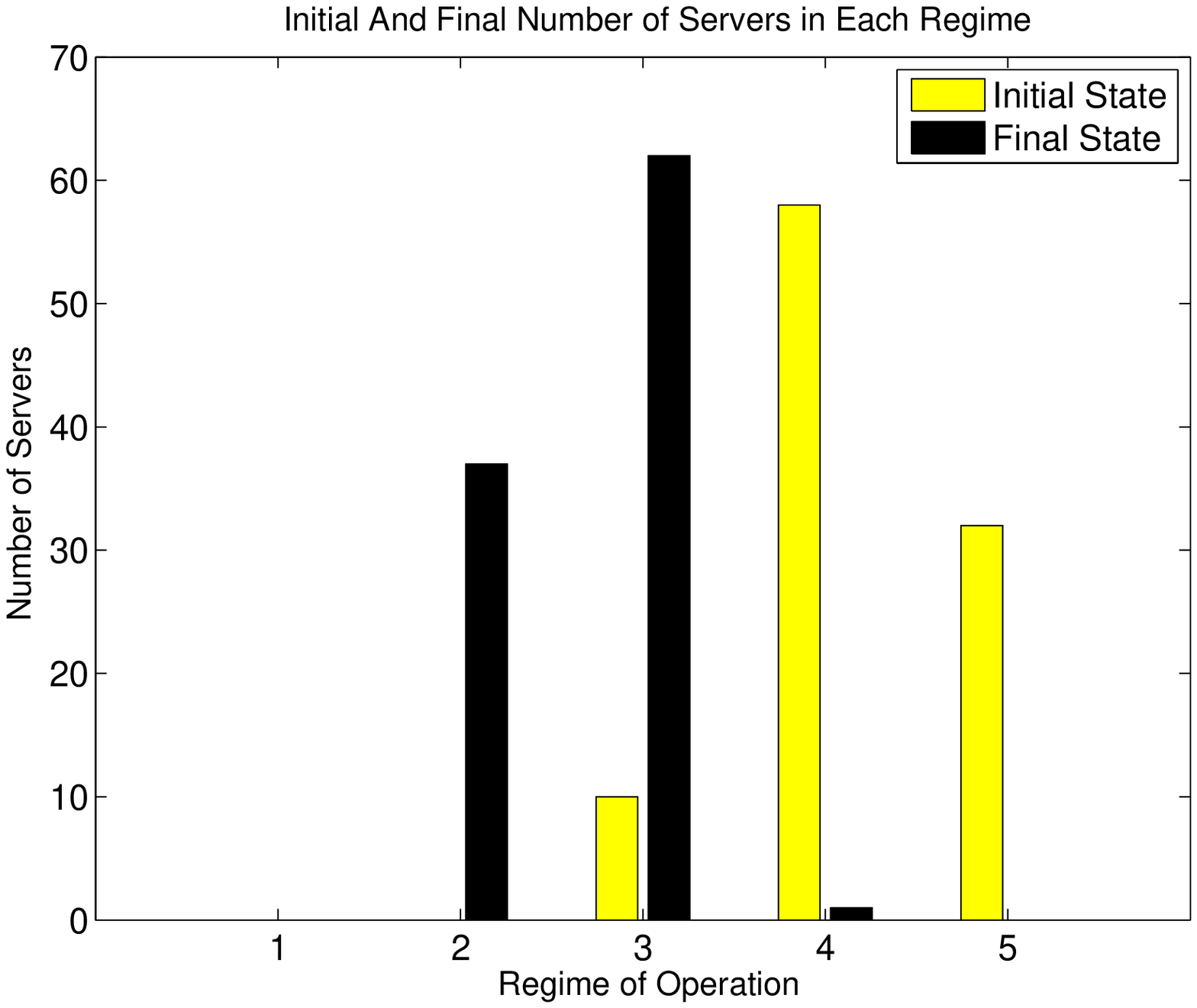}\\
~~~~(a) Cluster size: $10^{2}$. Average load: 30\%~~~~~~~~~~~~~~~~~~~~~(b) Cluster size $10^{2}$. Average load: -70\%\\
\includegraphics[width=9cm]{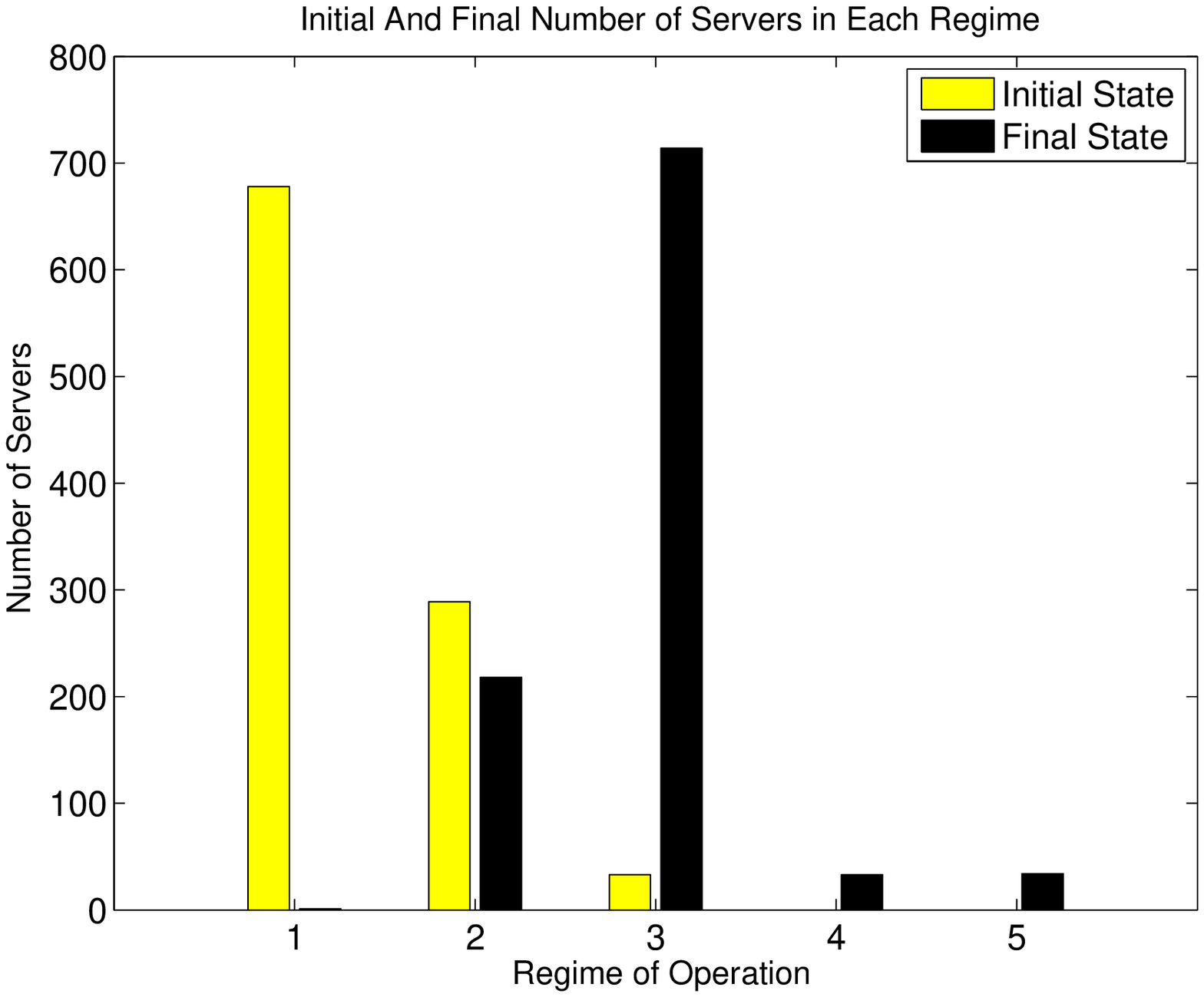}
\includegraphics[width=9cm]{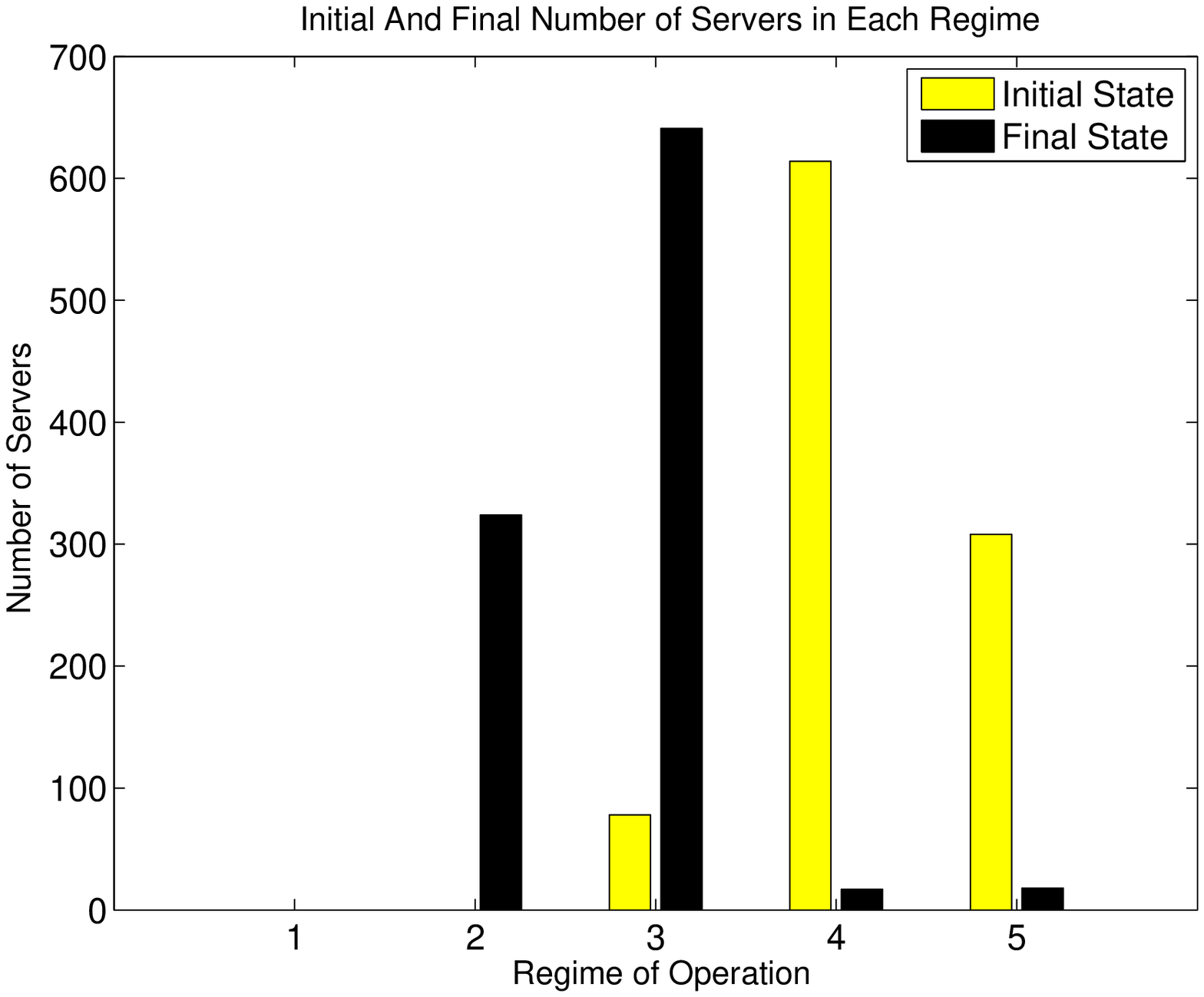}\\
~~~~(a) Cluster size: $10^{3}$. Average load: 30\%~~~~~~~~~~~~~~~~~~~~~(b) Cluster size $10^{3}$. Average load: -70\%\\
\includegraphics[width=9cm]{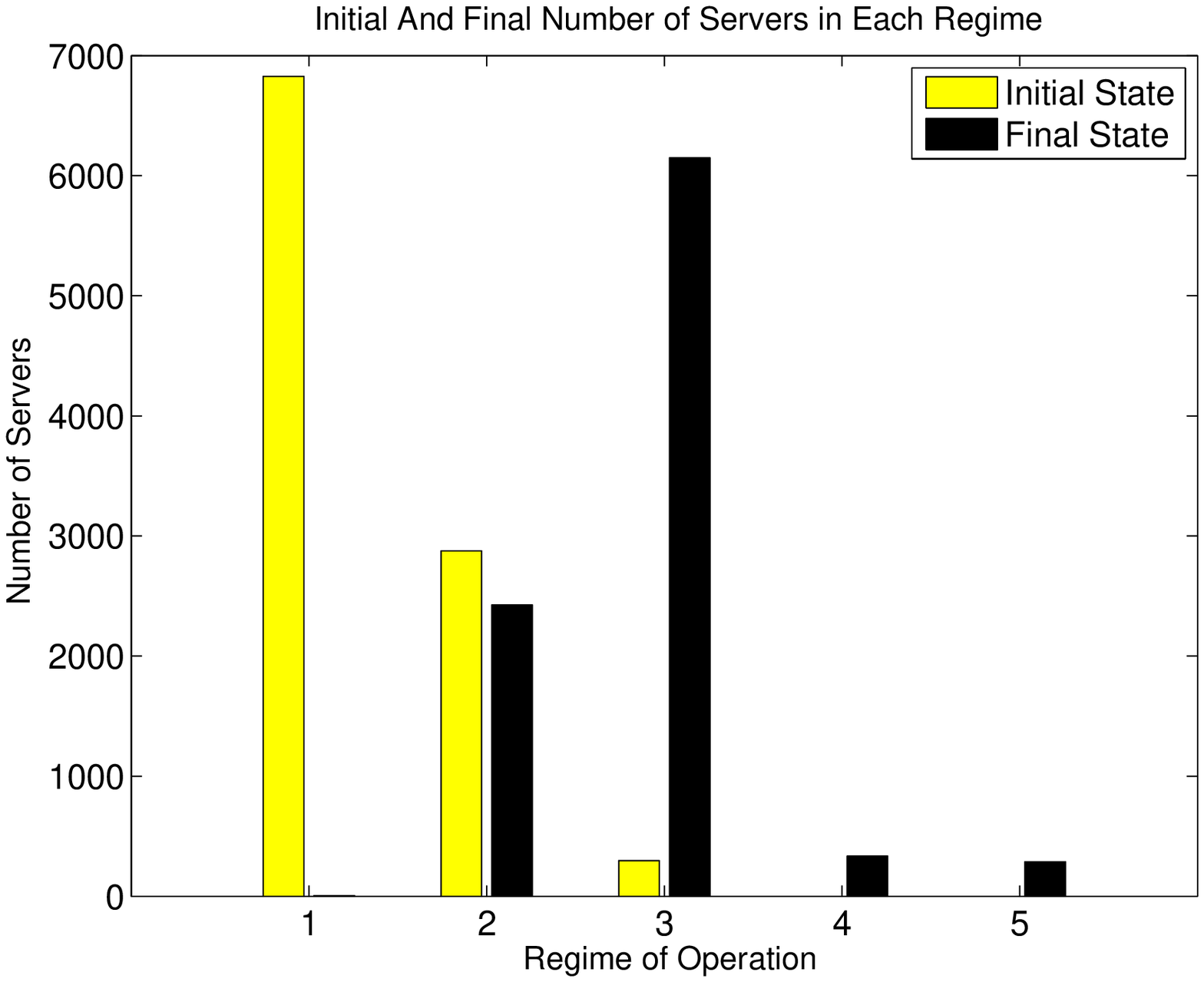}
\includegraphics[width=9cm]{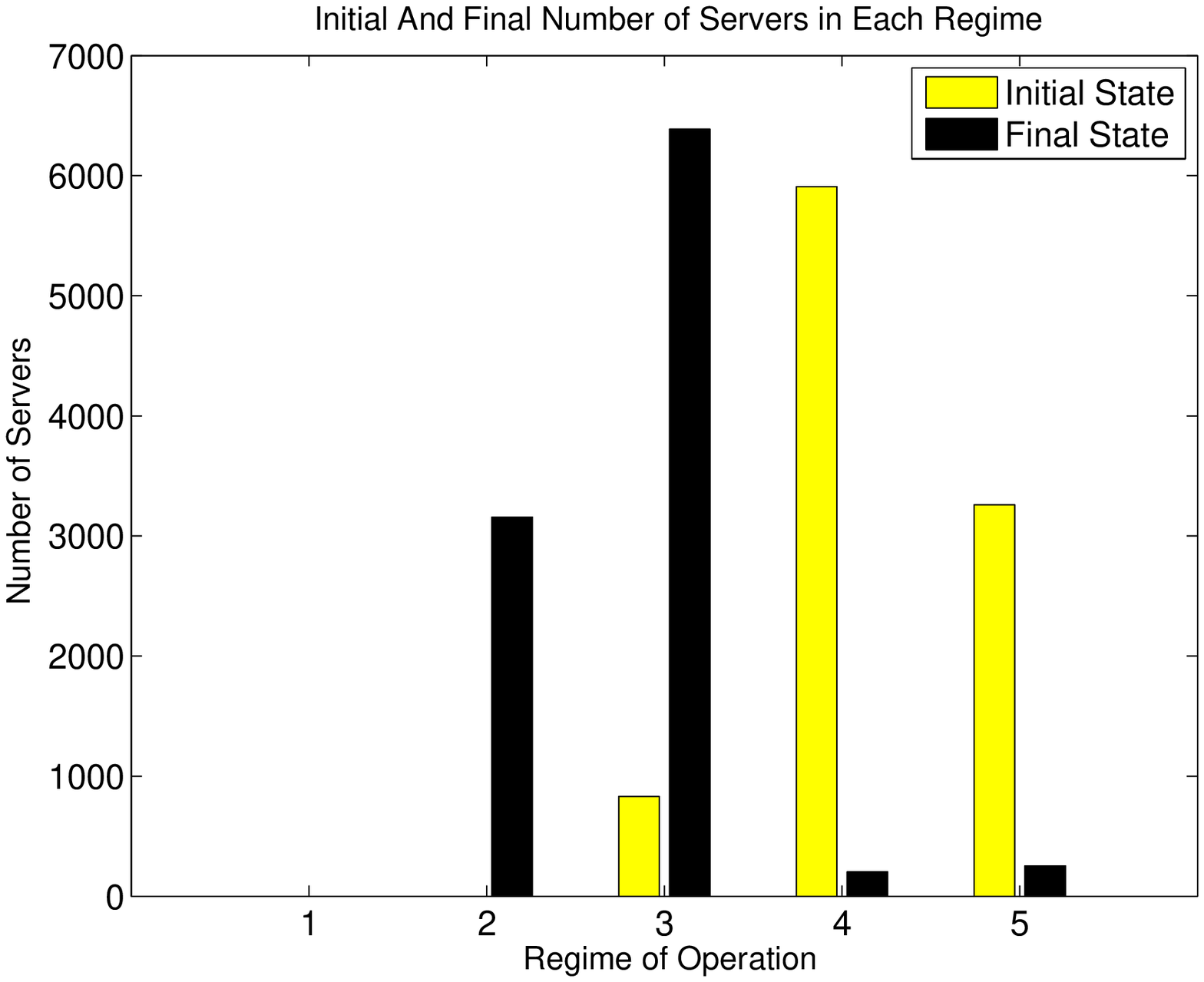}\\
~~~~(a) Cluster size: $10^{4}$. Average load: 30\%~~~~~~~~~~~~~~~~~~~~~(b) Cluster size $10^{4}$. Average load: -70\%\\
\end{center}
\caption{The effect of average server load on the distribution of the servers in the five operating regimes, $\mathcal{R}_{1}$, $\mathcal{R}_{2}$, $\mathcal{R}_{3}$, $\mathcal{R}_{4}$, and $\mathcal{R}_{5}$, before and after energy optimization and load balancing. Average load: (a), (c), (e) - 30\%; (b), (d), (f) - 70\%. The cluster size: (a) and (b) - $10^2$; (c) and (d) - $10^3$; (e) and (f) - $10^4$.}
\label{ClusterSize}
\end{figure*}

\smallskip

\noindent (i) Low average load - an initial load uniformly distributed in the interval $20-40\%$ of the server capacity. Figures \ref{ClusterSize} (a), (c), and (e) show the distribution of the number of servers in the five operating regions for clusters with $10^{2}, 10^{3}$ and $10^{4}$ servers, respectively, before and after load balancing. As expected, when average load $30\%$ of the server capacity, the initial server distribution is concentrated in operating regions at the left and in the optimal region $\mathcal{R}_{3}$. After load balancing, the majority of the servers operate within the boundaries of the optimal and the two suboptimal regimes, and almost $4\%$ in the undesirable regimes.

\smallskip

(ii) High average load - initial server load uniformly distributed in the $60-80\%$ of the server capacity. Figures \ref{ClusterSize} (b), (d), and (f) show the distribution of the number of servers in the five operating regions for clusters with $10^{2},10^{3}$ and $10^{4}$ servers, respectively, before and after load balancing.  In this case the average load is $70\%$ of the server capacity and the initial server distribution is concentrated in operating regions at the right of and in the optimal region. After load balancing, the majority of the servers operate within the boundaries of the optimal and the two suboptimal regimes, and almost $4\%$ in the undesirable regimes.

\begin{figure*}[!ht]
\begin{center}
\includegraphics[width=9cm]{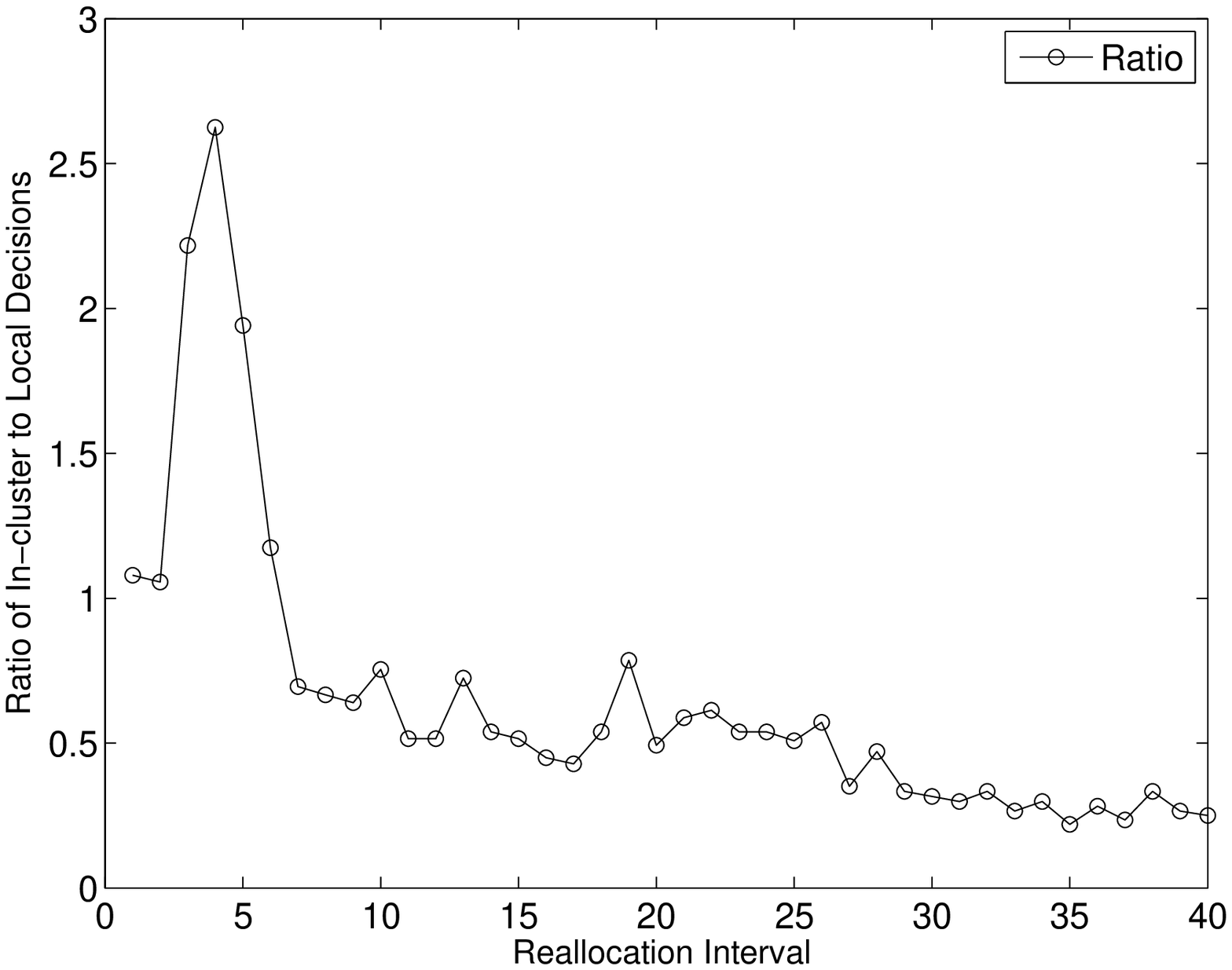}
\includegraphics[width=9cm]{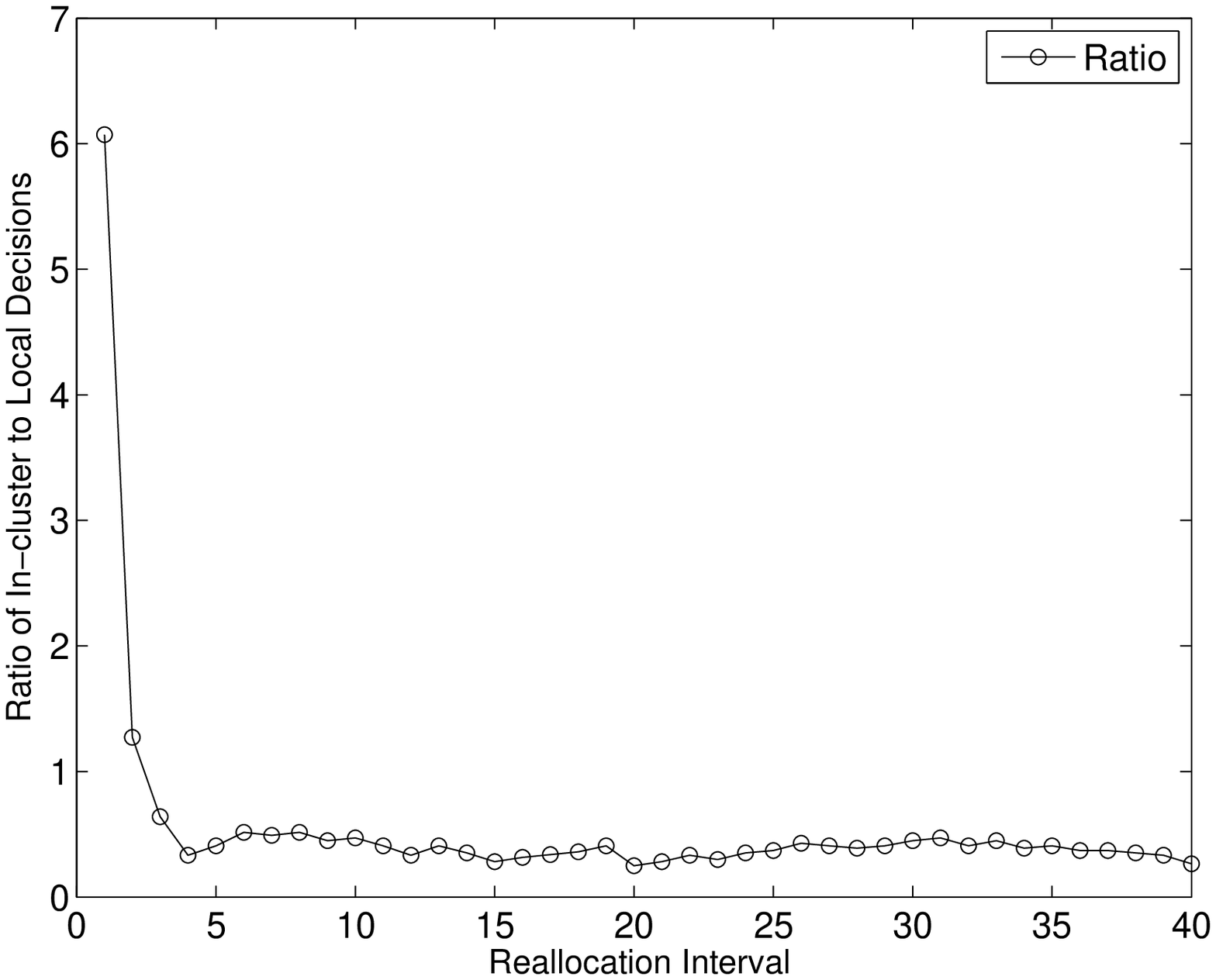}\\
(a)~~~~~~~~~~~~~~~~~~~~~~~~~~~~~~~~~~~~~~~~~~~~~~~~~~~~~~~~~~~~~~~~~~~(b)\\
\includegraphics[width=9cm]{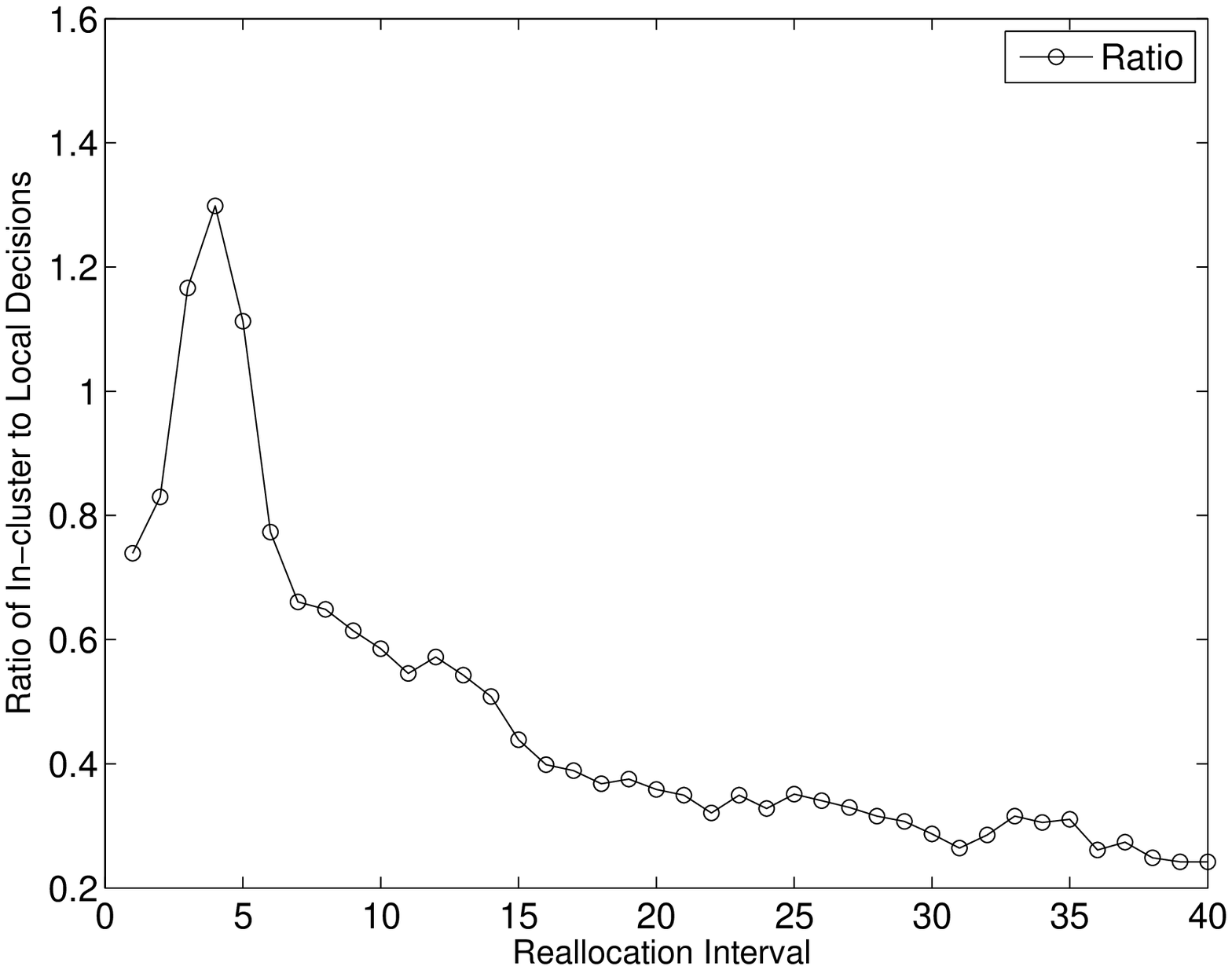}
\includegraphics[width=9cm]{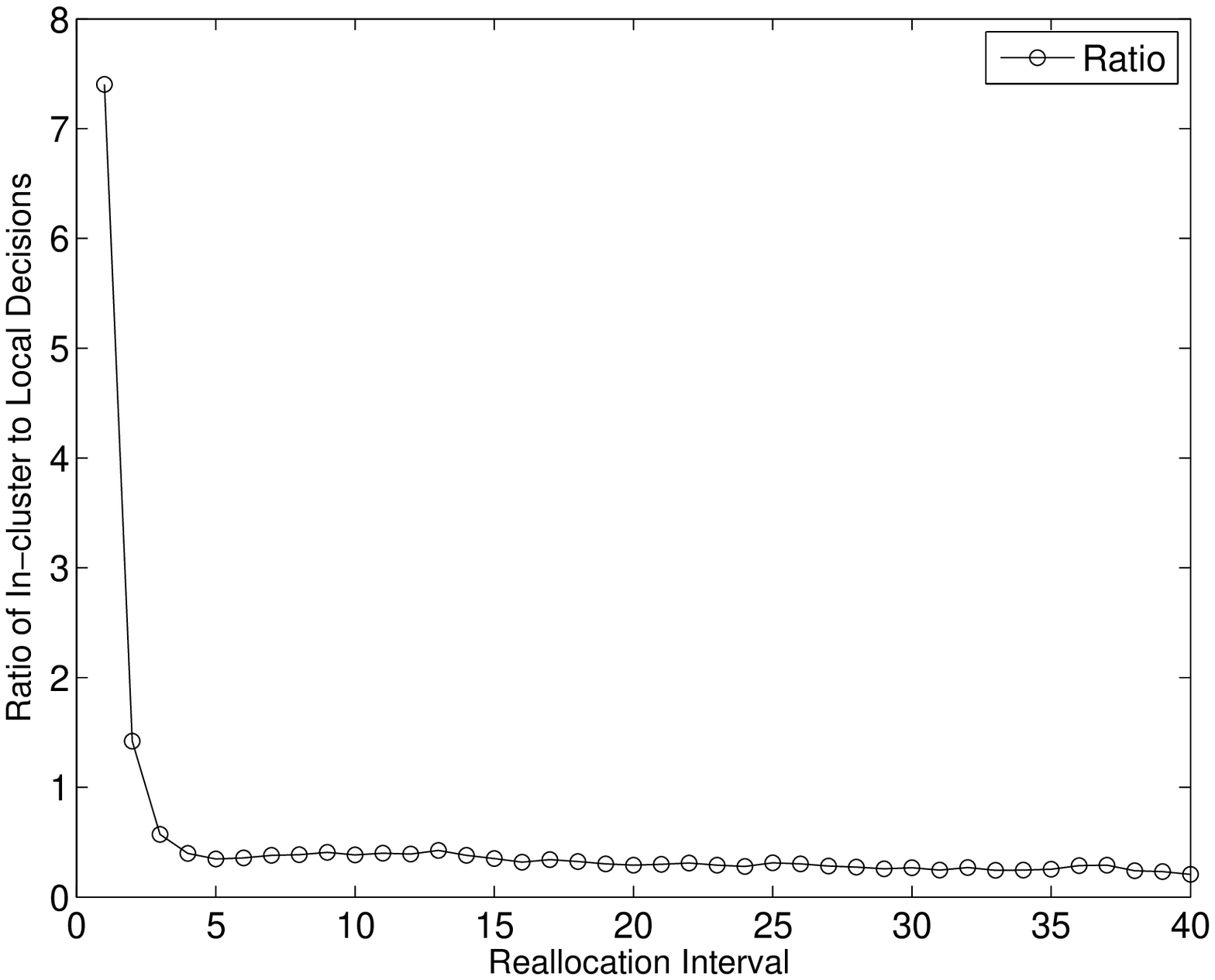}\\
(c)~~~~~~~~~~~~~~~~~~~~~~~~~~~~~~~~~~~~~~~~~~~~~~~~~~~~~~~~~~~~~~~~~~~(d)\\
\includegraphics[width=9cm]{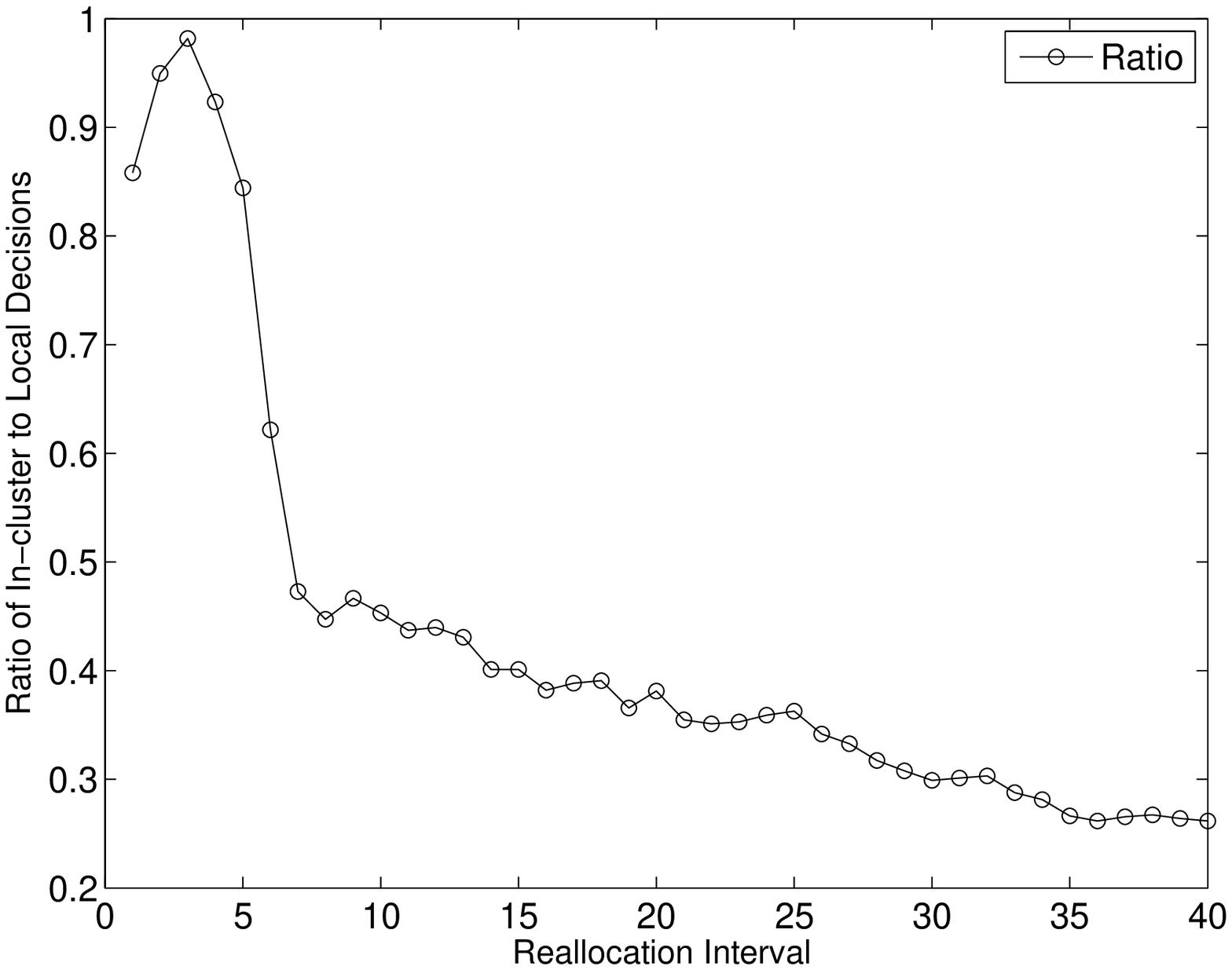}
\includegraphics[width=9cm]{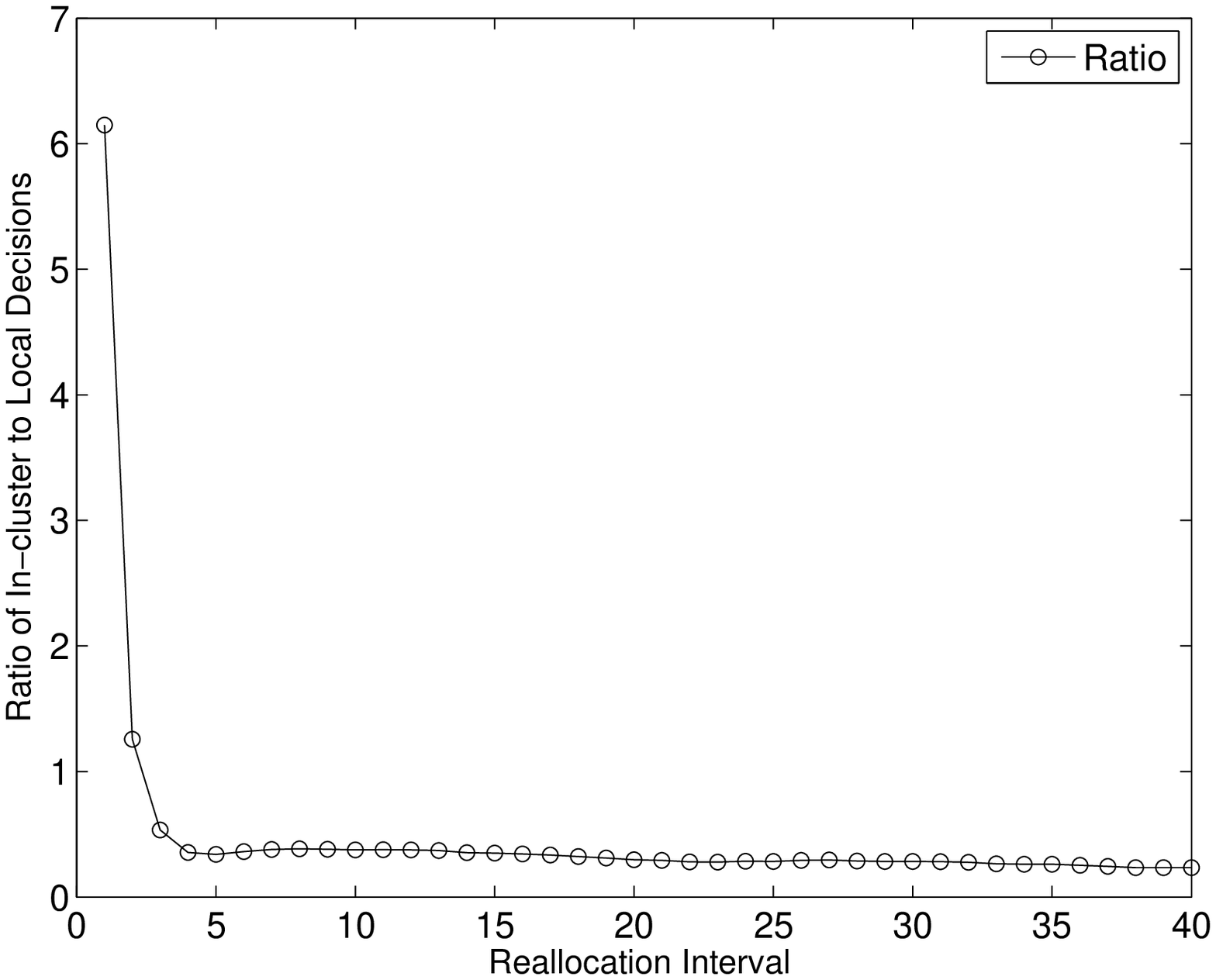}\\
(e)~~~~~~~~~~~~~~~~~~~~~~~~~~~~~~~~~~~~~~~~~~~~~~~~~~~~~~~~~~~~~~~~~~~(f)\\
\end{center}
\caption{Time series of in-cluster to local decisions ratios for $40$ reallocation intervals.  Average load: (a), (c), (e) - 30\%; (b), (d), (f) - 70\%. The cluster size: (a) and (b) - $10^2$; (c) and (d) - $10^3$; (e) and (f) - $10^4$. For low average load (a), (c) and (e) low-cost local decisions become dominant after about 20 reallocation intervals. For high average load (b), (d), and (f) low-cost local decisions become dominant after some 5 reallocation intervals.}
\label{RatioPlots}
\end{figure*}

\begin{table*}[!ht]

\caption{In-cluster to local decisions ratios for $30\%$ and $70\%$ average server load for three cluster sizes, $10^{2}$, $10^{3}$, and $10^{4}$.}
\label{LoadTable}
\begin{center}
\begin{tabular} {|c|c|c|c|c|c|}
\hline
Plot & Cluster & Average  & \# of servers  & Average  & Standard  \\
     & sizes   & load     & in sleep state &  ratio   & deviation \\       
     \hline \hline
(a)  & $10^2$  & 30\%     & 0              &  0.6490  &  0.5229  \\
(b)  & $10^2$  & 70\%     & 0              &  0.5540  &  0.9088  \\ 
\hline
(c)  & $10^3$  & 30\%     & 8              &  0.4739  &  0.2602  \\
(d)  & $10^3$  & 70\%     & 0              &  0.5248  &  1.1311  \\ 
\hline
(e)  & $10^4$  & 30\%     & 796            &  0.4294  &  0.1998  \\
(f)  & $10^4$  & 70\%     & 0              &  0.4843  &  0.9323  \\ 
\hline
\end{tabular}
\end{center}
\end{table*}

{\bf High-cost versus low-cost application scaling.} Elasticity is one of the main attraction of cloud computing; cloud elasticity allows application to seamlessly scale up and down.  

In the next set of simulation experiments we investigate horizontal and vertical application scaling for non-homogeneous clusters. {\it Horizontal scaling} requires the creation of additional VMs to run the application  on lightly loaded servers. Horizontal scaling  incurs higher costs for load balancing than vertical scaling. The higher costs are due to communication with the leader to identify the potential targets and then to transport the VM image to one or more of them. {\it Vertical scaling} allows VM running an application  to acquire additional resources from the local server; vertical scaling has lower costs, but it is only feasible if the server has sufficient free capacity.

We conduct six experiments for three cluster sizes, $10^{2}$, $10^{3}$, and $10^{4}$ and two different initial loads for each of them, $30\%$ and $70\%$ average server loads. We study the evolution of a cluster for some 40 reallocation intervals. We are interested in the average ratio of high-cost in cluster horizontal scaling to low-cost local vertical scaling and in the standard deviation of this ratio. 

Numerical results of these experiments are summarized  in Table~\ref{LoadTable}. For low average load Figures \ref{RatioPlots} (a), (c) and (e) show that low-cost local decisions become dominant after about 20 reallocation intervals. For high average load Figures \ref{RatioPlots} (b), (d), and (f) low-cost local decisions become dominant after some 5 reallocation intervals. As expected for low-average load after load balancing a number of servers are switched to the $C_{3}$ sleep state; this number increases from $0$ to $8$ and then to $796$ when the clusters size increases from  $10^{2}$ to $10^{3}$ and then to $10^{4}$ servers.

The average ratio of high to low cost scaling for the 40 reallocation intervals is in the $0.42$ to $0.65$ range, and decreases as the cluster size increases but has a large standard deviation due to the large variations for the first reallocation intervals. As the system stabilizes this ratio tends to have lower values.

\section{Summary}
\label{Conclusions}
\medskip

The average server utilization in large data-centers is $18\%$ \cite{Snyder10}. When idle the servers of a data center use more than half the power they use at full load. The alternative to the wasteful resource management policy when the servers are {\it always on}, regardless of their load, is to develop {\it energy-aware load balancing} policies. Such policies combine {\it dynamic power management} with load balancing. 

There are ample opportunities to reduce the energy necessary to power the servers of a large-scale data center and shrink the carbon footprint of cloud computing activities, even though this is only a fraction of the total energy required by the ever increasing appetite for computing and storage services. To optimize the resource management of large farms of servers we redefine the concept of load balancing and exploit the technological advances and the power management functions of individual servers. In the process of balancing the load we concentrate it on a subset of servers  and, whenever possible, switch the rest of the servers to a sleep state.

From the large number of questions posed by energy-aware load balancing policies we discuss only the energy costs for migrating a VM when we decide to either switch a server to a sleep state or force it to operate within the boundaries of an energy optimal regime. The policies analyzed in this paper aim to keep the servers of a cluster within the boundaries of the optimal operating regime. After migrating the VMs to  other servers identified by the cluster leader, a lightly loaded servers is switched to one of the sleep states. 

There are multiple sleep states; the higher the state number, the larger the energy saved, and the longer the time for the CPU to return to the state $C_{0}$ which corresponds to a  fully operational system. For simplicity we chose only two sleep states $C_{3}$ and $C_{6}$ in the simulation. If the overall load of the cluster is more than $60\%$ of the cluster capacity we do not switch any server to a $C_{6}$ state because in the next future the probability that the system will require additional computing cycles is high. Switching from the $C_{6}$ state to $C_{0}$ requires more energy and takes more time. On the other hand, when the total cluster load is less than $60\%$ of its capacity we switch to $C_{6}$ because it is so unlikely that for the next interval and the interval after that system needs extra computational unit.

The simulation results reported in Section \ref{SimulationExperiments} show that the load balancing algorithms are effective and that low-cost vertical scaling occurs even when a cluster operates under a heavy load. The larger the cluster size the lower the ratio of high-cost in-cluster versus low-cost local decisions.

The QoS requirements for the three cloud delivery models are different thus, the mechanisms to implement a cloud resource management policy based on this idea should be different. To guarantee real-time performance or a short response time, the servers supporting SaaS applications such as data-streaming or on-line transaction processing (OLTEP) may be required to operate within the boundaries of a sub-optimal region in terms of energy consumption.

There are cases when the instantaneous demand for resources cannot be accurately predicted and system are forced to operate in a non-optimal region before additional systems can be switched from a sleep state to an active one.   Typically, PaaS applications run for extended periods of time and the smallest set of serves operating at an optimal power level to guarantee the required turnaround time can be determined accurately. 

This is also true for many IaaS applications in the area of computational science and engineering. There is always a price to pay for an additional functionality of a system, so the future work should evaluate the overhead and the limitations of the algorithms required by these mechanisms.


\begin{thebibliography}{11}

\bibitem{Abts10}
D.~Abts.
\newblock ``The Cray XT4 and Seastar 3-D torus interconnect.''
\newblock {\it Encyclopedia of Parallel Computing, Part 3,} Ed. David Padua, pp. 470--477, Springer, 2011.


\bibitem{Abts10a}
D.~Abts, M.~R.~Marty, P.~M.~Wells, P.~Klausler, and H.~Liu.
\newblock ``Energy proportional datacenter networks.''
\newblock {\it ACM IEEE Int. Symp. on Comp. Arch. (ISCA'10)}, pp. 338--347, 2010.



\bibitem{Ardagna11}
D.~Ardagna, B.~Panicucci, M.~Trubian, and L.~Zhang.
\newblock ``Energy-aware autonomic resource allocation in multi-tier virtualized environments.''
\newblock {\it IEEE Trans. on Services Computing,} {\bf 5}(1):2--19, 2012.

\bibitem{Baliga11}
J. Baliga, R.W.A. Ayre, K. Hinton, and R.S. Tucker.
\newblock ``Green cloud computing: balancing energy in processing, storage, and transport.''
\newblock {\it Proc. IEEE}, {\bf 99}(1):149-167, 2011.


\bibitem{Barroso07}
L. A.~Barroso and U.~H\"ozle.
\newblock ``The case for energy-proportional computing.''
\newblock {\it IEEE Computer,} {\bf 40}(12):33--37, 2007.

\bibitem{Blackburn10}
M. Blackburn and A. Hawkins.
\newblock ``Unused server survey\break results analysis.''
\newblock {\it www.thegreengrid.org/media/White\break Papers/Unused\%20Server\%20Study\_WP\_101910\_v1.\break
ashx?lang=en} (Accessed on December 6, 2013).

\bibitem{Bodik09}
P. Bodik, R. Griffith, C. Sutton, A. Fox, M Jordan, and D. Patterson.
\newblock ``Statistical machine learning makes automatic control practical for Internet datacenters.''
\newblock {\it Proc. Conf. on Hot Topics in Cloud Computing.}, pp. 1-8, 2009.



\bibitem{Gandhi11}
A. Gandhi, and M. Harchol-Balter.
\newblock ``How data center size impacts the effectiveness of dynamic power management.''
\newblock {\it Proc. 49th Annual Allerton Conference on Communication, Control, and Computing, Urbana-Champaign}, pp. 1864--1869, 2011.


\bibitem{Gandhi12}
A. Ghandi, M. Harchol-Balter, R. Raghunathan, and M. A. Kozuch.
\newblock ``Autoscale: dynamic, robust capacity management for multi-tier data centers.''
\newblock {\it ACM Trans. on Computer Systems}, {\bf 30}(4):1--26, 2012.

\bibitem{Google13}
Google.
\newblock ``Google's green computing: efficiency at scale.''
\newblock {\it http://static.googleusercontent.com/external\_content/\break untrusted\_dlcp/www.google.com/en/us/green/pdfs/google\break
-green-computing.pdf} (Accessed on August 29, 2013).


\bibitem{Hasebe10}
K.~Hasebe, T.~Niwa, A.~Sugiki, and K.~Kato.
\newblock ``Power-saving in large-scale storage systems with data migration.''
\newblock {\it Proc IEEE 2nd Int. Conf. on Cloud Computing Technology and Science,}
pp. 266--273, 2010.

\bibitem{ACPI11}
Hewlet-Packard/Intel/Microsoft/Phoenix/Toshiba.
\newblock ``Advanced configuration and power interface specifications, revision 5.0''
\newblock {\it http://www.acpi.info/\break
DOWNLOADS/ACPIspec50.pdf}, 2011. (Accessed on November 10, 2013).

\bibitem{Koomey07}
J. G. Koomney.
\newblock ``Estimating total power consumtion by servers in the US and world.''
\newblock {\it http://hightech.lbl.gov/documents/data\_centers\break svrpwrusecompletefinal.pdf} (accessed on May 11, 2013).



\bibitem{LeSueur10}
E. Le Sueur and G. Heiser.
\newblock ``Dynamic voltage and frequency scaling: the laws of diminishing returns.''
\newblock {\it Proc. Workshop on Power Aware Computing and Systems, HotPower'10}, pp. 2--5, 2010.


\bibitem{Marinescu13}
D. C. Marinescu.
\newblock ``Cloud Computing; Theory and Practice.''
\newblock {\it Morgan Kaufmann}, 2013.

\bibitem{Mazzucco10}
M.~Mazzucco, D.~Dyachuk, and R.~Deters.
\newblock ``Maximizing cloud providers revenues via energy aware allocation policies.''
\newblock {\it Proc. IEEE 3rd Int. Conf. on Cloud Computing}, pp. 131--138, 2010.

\bibitem{Mills13}
M. P. Mills
\newblock ``An overview of the electricity used by the global digital ecosystem.''
\newblock {\it http://www.tech-pundit.com/wp-content/uploads/2013/07/Cloud\_Begins\break
\_With\_Coal.pdf}, 2013. (Accessed on September 22, 2013)


\bibitem{NRDC12}
NRDC and WSP 2012.
\newblock ``The carbon emissions of server computing for small-to medium-sized organization - a performance study of on-premise vs. the cloud.''
\newblock {\it http://www.wspenvironmental.com/media/docs/\break ourlocations/usa/NRDC-WSP\_Cloud\_Computing.pdf} October 2012 (Accessed on November 10, 2013).


\bibitem{Paya13}
A. Paya and D. C. Marinescu.
\newblock ``Energy-aware application scaling on a cloud.''
\newblock http://arxiv.org/pdf/1307.3306v1.pdf, July 2013.

\bibitem{Preist10}
C.~Preist and P.~Shabajee.
\newblock ``Energy use in the media cloud.''
\newblock {\it Proc IEEE 2nd Int. Conf. on Cloud Computing Technology and Science,}
pp. 581--586, 2010.

\bibitem{Snyder10}
B. Snyder.
\newblock ``Server virtualization has stalled, despite the hype.''
\newblock {\it http://www.infoworld.com/print/146901} (Accessed on December 6, 2013)

\bibitem{Urgaonkar05}
B. Urgaonkar and C. Chandra.
\newblock ``Dynamic provisioning of multi-tier Internet applications.''
\newblock {\it Proc. 2nd Int. Conf, on Automatic Computing}, pp. 217–-228, 2005.

\bibitem{Van10}
H.~N.~Van, F.~D.~Tran, and J.-M.~Menaud.
\newblock ``Performance and power management for cloud infrastructures.''
\newblock {\it Proc. IEEE 3rd Int. Conf. on Cloud Computing}, pp. 329--336, 2010.


\bibitem{Verma09}
A. Verma, G. Dasgupta, T. K. Nayak, P. DE, and R. Kothari.
\newblock ``Server workload analysis for power minimization using consolidation.''
\newblock {\it Proc. USENIX'09 Conf.}, pp.28--28, 2009.


\bibitem{Vrbsky10}
S.~V.~Vrbsky, M.~Lei, K.~Smith, and J.~Byrd.
\newblock ``Data replication and power consumption in data grids.''
\newblock {\it Proc IEEE 2nd Int. Conf. on Cloud Computing Technology and Science,}
pp. 288--295, 2010.

\end{thebibliography}
\end{document}